\documentclass[preprint]{elsarticle}

\usepackage{amsmath}
\usepackage{epstopdf}
\usepackage{epsfig}

\begin{document}
\begin{frontmatter}

\title{Scintillation detectors of Alborz-I experiment}

\author[mymainaddress,mysecondaryaddress]{Y.~Pezeshkian}

\author[mymainaddress,mysecondaryaddress]{M.~Bahmanabadi\corref{mycorrespondingauthor}}
\cortext[mycorrespondingauthor]{Corresponding author}
\ead{bahmanabadi@sharif.edu}

\author[mysecondaryaddress]{M.~Abbasian Motlagh}

\author[mysecondaryaddress]{M.~Rezaie}

\address[mymainaddress]{Department of Physics, Sharif University of Technology,
P.O.Box 11155-9161, Tehran, Iran}
\address[mysecondaryaddress]{Alborz Observatory, Sharif University of Technology,
P.O.Box 11155-9161, Tehran, Iran}

\begin{abstract}
A new air shower experiment of the Alborz Observatory, Alborz-I,
located at the Sharif University of Technology, Iran, will
be constructed in near future. An area of about  30$\times$40
m$^{2}$ will be covered by 20 plastic scintillation detectors (each
with an area of 50$\times$50 cm$^{2}$). A series of experiments have
been performed to optimize the height of light enclosures of the
detectors for this array and the results have been compared to an
extended code simulation of these detectors. Operational parameters
of the detector obtained by this code are cross checked by Geant4
simulation. There is a good agreement between extended-code and
Geant4 simulations. We also present further discussions on the 
detector characteristics, which can be applicable for all
scintillation detectors with a similar configuration.
\end{abstract}

\begin{keyword}
Cosmic rays, Scintillation detectors, Ground based arrays
\PACS 95.55.Vj, 95.85.Ry, 95.75.-z, 29.40.Mc
\end{keyword}

\end{frontmatter}

\section{Introduction}
Nowadays scintillators are among important detectors used in 
many cosmic ray and high energy physics experiments.
For instance, they are a major part of detectors in balloon 
experiments for the detection of cosmic rays; such as TIGER, CREAM 
and TRACER \cite{2012-Seo-AP}. Liquid scintillators are used in 
several neutrino experiments, such as Borexino and KamLAND and 
in the next generation neutrino experiment LENA \cite{2011-Wurm-AP}.

Scintillators at ground based arrays are used in different
configurations. Liquid scintillators with pyramidal light enclosure
and PhotoMultiplier Tube (PMT) at the vertex are used in KASCADE
array \cite{2003-Ulrich-NIMA}. In many experiments the
plastic scintillators are connected to PMTs with optical fibers as
light guide and wavelength shifter (WLS), e.g., Telescope array
\cite{2012-Nonaka-NIMA}, KASCADE-Grande \cite{2010-DiPierro-NIMA},
and LHAASO \cite{2014-Cui-AP}. Detectors in experiments of 
Tibet \cite{2011-Amenomori-ASTRA} and GRAPES3 \cite{2005-Gupta-NIMA} 
enclose plastic scintillators in pyramidal light enclosures. 
Both of these experiments also have scintillation detectors with optical
fibers\cite{2011-Amenomori-ASTRA, 2009-Gupta-AP}. Time resolution of
scintillators with light enclosure is better than scintillators with
WLS fiber readouts, since WLS has a large contribution in time
resolution due to the decay time of WLS. On the other hand,
scintillators with WLS provide more uniformity compared to light
enclosures. Choosing detectors' configuration depends on the
requirements of the experiment and the budget restrictions.

The Alborz air shower observatory at the Sharif University of Technology
($35^{\circ}43^{'}$N, $51^{\circ}20^{'}$E, 1200 m a.s.l, $890\, {\rm
g}\, {\rm cm}^{-2}$) will have 30 installed detectors (20 scintillators 
and 10 Cherenkov detectors). The project is planned to be accomplished 
in 2 stages. At first stage (Alborz-I), the scintillation detectors 
with the area of $50\times50\,{\rm cm}^{2}$ will be placed on 
the rooftop of the highest building of the university campus. 
Detectors will be arranged in a square lattice with a side of 
$\sim 7\, {\rm m}$ (Fig.1(a)). Required electronics have already been 
prepared and primary tests will be accomplished in future. The aim
of Alborz-I is to study extensive air showers (EAS) with energies
around the knee region of the cosmic rays spectrum. Later, by adding
more detectors, energy regions above the knee will be investigated.
Primary studies, like detailed simulation of array configuration and
distance between detectors, are under investigation. Cherenkov detectors 
will be added at second stage. Dimensions and some properties of Cherenkov 
detectors are optimized and the results were published in
\cite{2012-Mortazavi-AP}. Detailed studies of scintillation detectors 
and related optimizations are subject of this paper.

In the former arrays of the Alborz observatory
\cite{2002-Bahman-EA, 2003-Bahman-EA, 2005-Khakian-A&A}, the size of
the scintillators was $100 \times 100\, {\rm cm}^2$ and the
performance of these scintillators had studied in \cite{1998-Bahman-EA}. 
In order to improve the accuracy of the array in determining directions 
of cosmic rays, the size of detectors is reduced to 
$50 \times 50\, {\rm cm}^2$. By resizing the scintillators, 
light enclosure heights should also be optimized to obtain suitable 
signals. Beside height optimization of the light enclosures, we will 
also present the further results on the performance of these detectors
including time response, uniformity, and effect of walls of the
light enclosure.

In order to find the best height for the light enclosures, three
different light enclosures are built and compared by experiment.
Experimental set-up and procedure is explained and corresponding
results are discussed in section 2. Section 3 with three 
subsections is dedicated to explain an extended-code simulation 
of Alborz-I scintillators. We describe details of the algorithm 
at the first subsection. In the second subsection, we present 
the simulation results including time spectrum of photons 
reaching PMT, the effect of light enclosure walls on the total 
number of captured photons, and other parameters of the detectors. 
At the third subsection, in order to cross check our extended 
code simulation, we compare time spectrum of photons reaching 
PMT obtained by Extended-code 
with a Geant4 simulation. we conclude that they have similar 
results for the same initial conditions. Some further discussions 
about time resolution of detectors are also presented in this 
subsection.

\section{Experimental setup and Results}

\begin{figure}
\begin{center}
\hspace{1cm}\includegraphics*[width=0.7\textwidth]{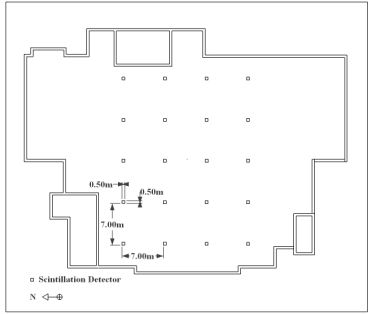}
\\
(a)
\\
\includegraphics*[width=0.40\textwidth]{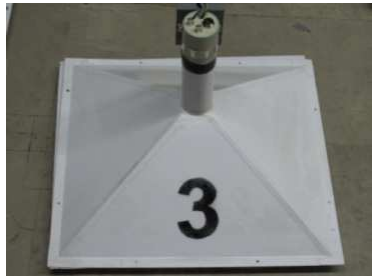}\hspace{0.5cm}\includegraphics*[width=0.31\textwidth]{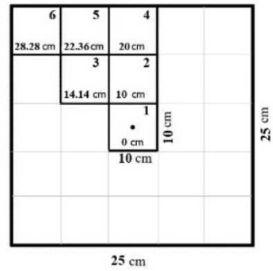}
\\
(b)\hspace{4.2cm}(c)
\includegraphics*[width=0.7\textwidth]{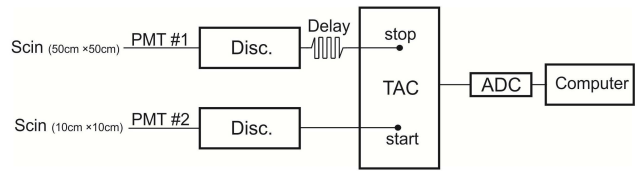}
\\
(d)
\end{center}
\caption{\label{fig:detector} (a) A schematic view of Alborz-I
array, (b) A sample of the Alborz-I scintillation detectors, (c) 6
swept cells of scintillator surface; the number in each cell is
distance from the center of the detector, and (d) electronic circuit
for data Read out.}
\end{figure}

The detectors of the Alborz I observatory are plane plastic 
scintillators (NE102A) which are enclosed in a pyramidal light 
enclosure with a 5 cm photomultiplier tube (PMT, 9813B) at the 
vertex of the light enclosure (Fig.~\ref{fig:detector}(b)). In order 
to optimize the height of the light enclosure, two parameters are 
important: total efficiency of the detector, and its uniformity. 
Uniformity requirement can be satisfied when the detection parameters 
(i.e., total number of events in the cells and their time resolutions) 
encounter little variation from center to the edge of the detector.

If the height of the enclosure is small, the PMT is very close to
the center of the scintillator (and consequently far from the
corners), therefore it is not possible to have an acceptable
uniformity. By increasing the height of the light enclosure,
uniformity of the detector from center (cell 1 in
Fig.~\ref{fig:detector}(c)) to corner (cell 6 in
Fig.~\ref{fig:detector}(c)) improves, but count rate of particles
(i.e., the efficiency) decreases. To determine the optimum height, three
light enclosures with heights of $10\, {\rm cm}$, $20\, {\rm cm}$,
and $30\, {\rm cm}$ were made and a series of experiments for each
of them were carried out.

The bottom surface of the scintillator is divided into 25 cells, each
of them with the area $10\times 10\, {\rm cm}^2$, as shown in
Fig.~\ref{fig:detector}. Due to symmetry, it suffices to choose only six
cells (cells 1-6 in Fig.~\ref{fig:detector}(c)), and sweep them
by a smaller scintillator ($10\times 10\times 0.5\, {\rm cm}^3$) and
accumulate the coincidence of the events for each cell and the small
scintillator. In all experiments, the time lag between signals of
these two scintillators is measured. Fig. 1(d) displays the scheme
of the electronics in more detail, including fast discriminator
(Disc., N413A), Time-to-Amplitude Converter (TAC, ORTEC566), and
Analogue-to-Digital converter (ADC, KIAN AFROUZ). Output of this
circuit is the time lag of the passage of a cosmic ray through both
detectors. Overall run time of the experiment for 3 heights and
6 cells was 3$\times$6$\times$22 h. Temperature fluctuation was
negligible during one month of the experiment, therefore, the effect of
temperature and pressure fluctuations on the cosmic ray flux is not
considered.

\begin{figure}[h]
\begin{center}
\includegraphics*[width=0.7\textwidth]{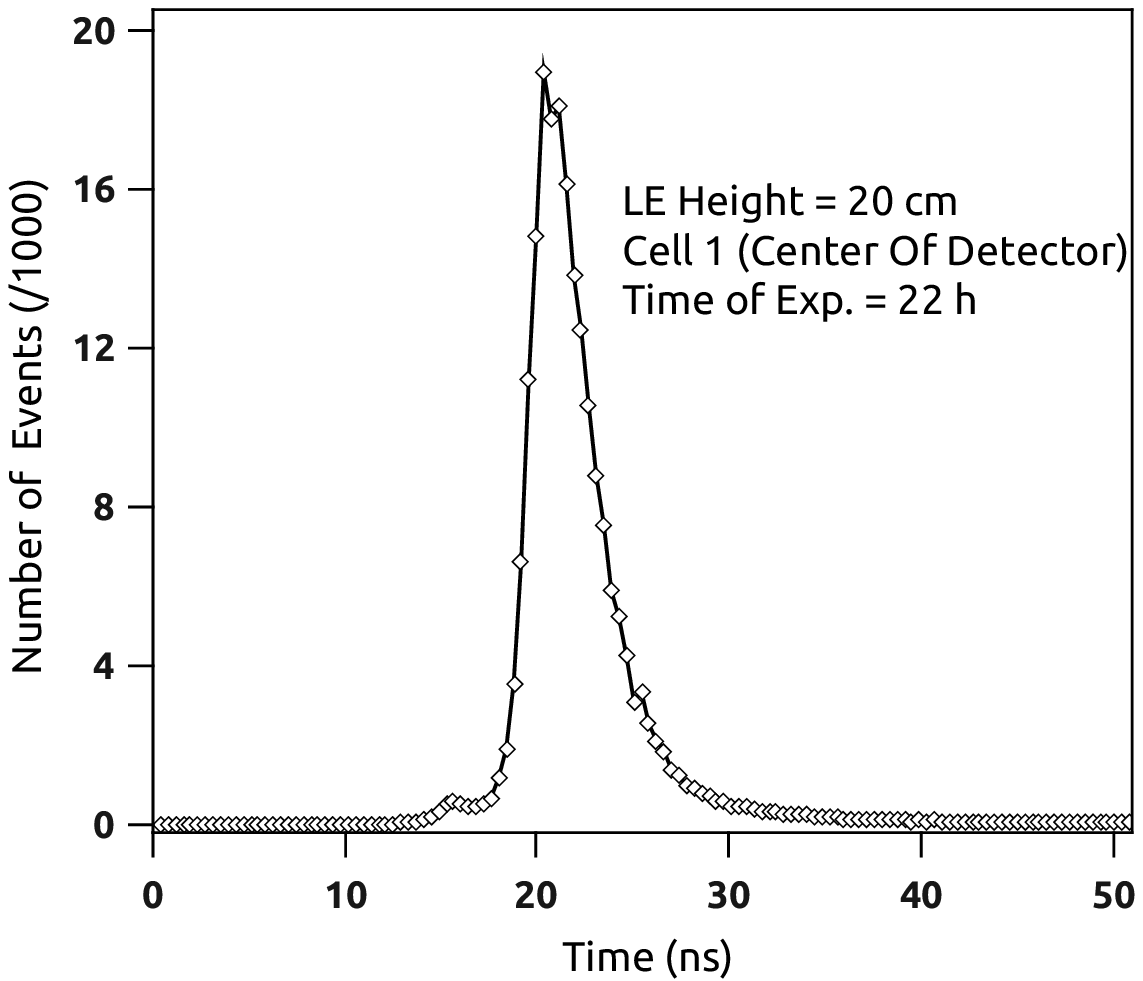}
\end{center}
\caption{\label{fig:timeLag} A sample of time lag distributions.}
\end{figure}

A sample of the time lag distribution of particles passing through
two detectors is shown in Fig.~\ref{fig:timeLag}. In this spectrum,
each count is a start-stop event recorded by the Time-to-Amplitude
Converter (TAC), which was set at a full scale of 200~ns, and
corresponds to the passage of a cosmic ray through both
scintillators in our experimental setup. The main differences in the
spectra are the number of total counts under the peak, and the width
of the peak. In computing the total counts under the peak for each
spectrum, very low background counts (obtained from channels far
from the peaks) have been subtracted. The width of each peak shows
the time resolution. In extensive air shower experiments, which
employ the large area scintillators and the light enclosures viewed
by a single PMT, the uncertainty in transit time measurement
produces an error in determining the arrival direction of
the total shower. This transit time error component is in addition
to the uncertainty in the transit position in the large area
scintillator. The width of each peak is interpreted as an inherent
timing error due to the light enclosures and the electronic devices.
The Full Width Half Maximum (FWHM) of each time lag distribution can
be obtained by fitting a Gaussian function to it
(Fig.~\ref{fig:timeLag}). The uniformity of each detector can be
evaluated by comparing the FWHM ($\Gamma$) and the efficiency
($\varepsilon$) values of its different cells.

\begin{figure}[h]
\begin{center}
\includegraphics*[width=0.7\textwidth]{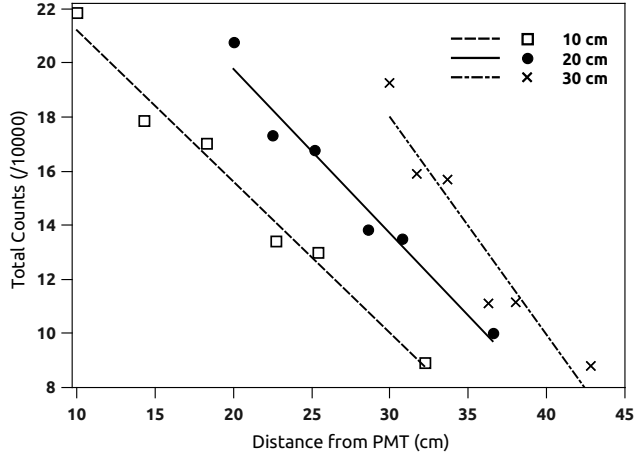}
\end{center}
\caption{\label{fig:Total counts} Total counts in each cell as a
function of distance between corresponding cell and PMT for 3 light
enclosures.}
\end{figure}

Fig.~\ref{fig:Total counts} shows the total counts in each cell for
heights of 10~cm, 20~cm, and 30~cm light enclosure. This figure
shows a considerable non-uniformity in total counts in different
cells.

\begin{figure}[h]
\begin{center}
\includegraphics*[width=0.7\textwidth]{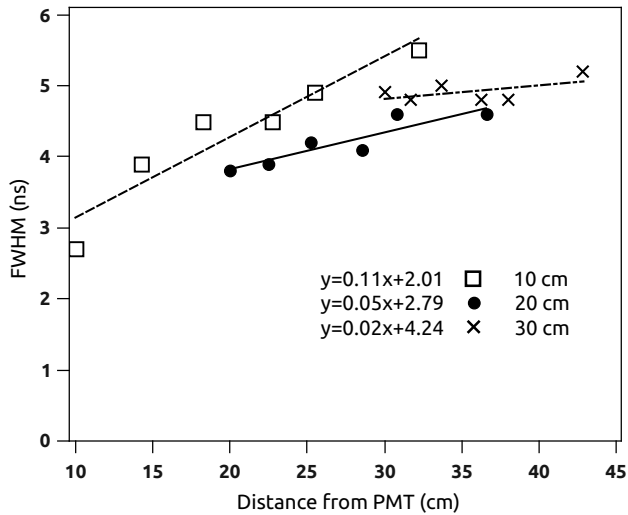}
\end{center}
\caption{\label{fig:fwhm} FWHM of each cell as a function of
distance between corresponding cell and PMT for 3 light enclosures.
Since the error bars are smaller than the data points, they are
omitted.}
\end{figure}

In Fig.~\ref{fig:fwhm}, we show the $\Gamma$ value of each cell as a
function of the cell distance from the PMT, for three heights.
Equations of the fitted lines are also shown in the figure and the
slope of each line is available in table 1. The statistical errors
are not visible in Fig.~\ref{fig:fwhm}, since they are smaller than
the data points, but the deviations of the points from the lines are
arisen due to asymmetry of the pyramidal light enclosure as a
systematic error. In the case of 10~cm light enclosure, slope of the
line is 0.11~ns/cm, which means that the FWHM of the time lag
distributions increases significantly from the center to the corner
of the detector(about 2.26~ns).

Mean value of FWHM for each height of the light enclosure can be
calculated using the following equation:
\begin{equation}
\bar{\Gamma}=\dfrac{1}{\sum_{j=1}^{6} m_{j}N_{i}} \sum_{j=1}^{6}
m_{i}N_{i}\Gamma_{i} ;
\end{equation}
where the summation is over the 6 swept cells
(Fig.~\ref{fig:detector}(c)), $N_{i}$ is the total number of events
in the cell $i$ (Fig.~\ref{fig:Total counts}) and $m_{i}$ shows
number of cells similar to the cell $i$ on the total surface of the
scintillator (that is $m_{1}=1$, $m_{5}=8$, and $m_{i} = 4$ for the
rest of the cells). It is clear that the lower value of
$\bar{\Gamma}$ indicates better time response of the detector.
Therefore, the 20 cm light enclosure with FWHM of 4.35 ns has the 
best time resolution (Table \ref{tab:table0}).

In addition to these two parameters ($\bar{\Gamma}$ and the slope of
$\Gamma$), the total number of events (i.e $\sum m_{i}N_{i}$), is
another criterion for comparing the different light enclosures. Since
the rate of cosmic rays is constant, one expects the number of
particles crossing surfaces with equal areas ($10\times 10\,{\rm
cm}^2$) at equal times (22~hours) to be the same. We assume the
number of particles passing through each cell ($N_{T}$) is the same
for all cells, namely, the flux of particles does not change during
one month. The efficiency of the cell $i$ is defined as
$\varepsilon_{i} = N_{i} / N_{T}$, where $N_{i}$ is the number of
recorded particles in the cell $i$. The mean efficiency of the
detector could be obtained as a function of the efficiency of cell
in which it is maximum (in our case the cell 1 of the 10~cm light
enclosure):

\begin{equation}
\bar{\varepsilon} = \dfrac{1}{\sum_{i=1}^{6} m_{i}} \sum_{i=1}^{6}
\dfrac{m_{i}N_{i}}{N_{1,10}}\varepsilon_{0} \times{100} ;
\end{equation}
where ${N_{1,10}}$ and ${\varepsilon_{0}}=\dfrac{N_{1,10}}{N_{T}}$
are number of events and efficiency of the cell 1 with the 10 cm
light enclosure, respectively. 

\begin{table}
\begin{center}
\begin{tabular}{|c|c|c|c|}
\hline
\rule{0pt}{10pt} Height(cm) & $\bar{\varepsilon}$(\%) & $ \bar{\Gamma} $(ns) & slope of $\Gamma$(ns~$cm^{-1}$)\\
\hline
 10 & (64.9 $\pm$ 1.6) $\varepsilon_{0}$ & 4.73 $\pm$ 0.09 & 0.11 \\
 20 & (66.0 $\pm$ 1.6) $\varepsilon_{0}$ & 4.35 $\pm$ 0.08 & 0.05 \\
 30 & (57.5 $\pm$ 1.4) $\varepsilon_{0}$ & 5.01 $\pm$ 0.11 & 0.02 \\
\hline
\end{tabular}
\caption{\label{tab:table0} $\bar{\varepsilon}$, $\bar{\Gamma}$ (ns)
and Slope of $\Gamma$ (ns~$cm^{-1}$) for various heights of the light
enclosures.}
\end{center}
\end{table}

Results of the experiment (Table \ref{tab:table0}) show that the
20~cm light enclosure has a better efficiency and its mean value of
FWHM ($\bar{\Gamma}$) is smaller compared to the 10~cm and 30~cm
light enclosures. If the uniformity is the main concern, the 30~cm light
enclosure should be chosen, however, because of its lower efficiency
and higher $\bar{\Gamma}$, this light enclosure is not suitable. Therefore
a 20~cm height is selected for the light enclosure of the detector.

\section{Simulation}
\subsection{Algorithm} \label{SS3.1}
A simulation of what happens in the detectors enables us to
understand the detection process in more detail. When a charged
particle passes through the detector, it loses some or all of its
kinetic energy in the scintillator. Small fraction of the lost
energy will be converted into fluorescent light and the remainder
will be dissipated nonradiatively, primarily in the form of lattice
vibrations and heat \cite{Knoll-2000}. The number of scintillation
photons can be calculated by using the light yield of the scintillator
(available in the literature). In the simulation we use the properties
of the scintillators of our experiment i.e., NE102A.

For the energy loss of a charged particle, $dE/dx$, in the
scintillator, we use the value $2\, {\rm MeV}\, {\rm cm}^{-1}$,
which is consistent with $\rho^{-1}\, dE/dx = 2\, {\rm MeV}\,/( {\rm
g} {\rm cm}^{-2}$) \cite{Grupen-2008}, and the density of NE102A,
$\rho = 1.032\, {\rm g}\, {\rm cm}^{-3}$ \cite{Leo-2008}.

Since light yield of NE102A is 10,000~photons~${\rm MeV}^{-1}$
\cite{Knoll-2000}, about 20,000 photons will be emitted through each
cm of the scintillator.  A scintillator does not emit all photons
simultaneously, hence the profile of the light pulse of the
scintillator should be considered. The distribution function of the
emission time of photons is given by:
\begin{equation}
 I = I_{0}(e^{-t/\tau_{1}}-e^{-t/\tau_{2}}) ;
\end{equation}
where $\tau_{1}$ and $\tau_{2}$ are equal to 2.4 ns and 0.6 ns, 
respectively \cite{Knoll-2000}.

For each photon, the emission time can be assigned by using Monte
Carlo method. In order to reduce run time of the program, emission
time of photons is calculated only for photons reaching PMT, after
finishing other steps of the simulation.

Each emitted photon chooses a stochastic direction and passes inside
the detector. It is annihilated by two ways: (1) absorption by the
inner surfaces of the light enclosure or its bottom surface (i.e., 
the surface directly beneath the large area scintillator), and 
(2) attenuation inside the plastic scintillator. The inner surface 
of the light enclosure is diffuse white, and its bottom surface is 
made of aluminium foil. By Diffuse Reflectance Spectroscopy (DRS) test,
their absorption coefficients are about 0.2 and 0.1, respectively. A
signal can be produced by photons reaching PMT.

The whole process can be summarized in the following algorithm:
\begin{enumerate}
\item Based on the path length of a particle inside the scintillator,
the number of created photons can be obtained.

\item Then, a random direction ($\theta$ and $\varphi$) is assigned
to each photon.

\item The photon path is followed until it hits the surface of the 
scintillator. If $\theta$ is greater than the critical angle, it will 
undergo total internal reflection, i.e., it will be mirrored inside the
scintillator. Otherwise, it will get out of the scintillator. By
using the geometry of the detector (pyramidal shape of the light
enclosure) and equations of the detector surfaces, distance between
start point and hit point of the photon with one of the light
enclosure surfaces is calculated.

\item At this stage, four different cases can happen:\\
    a. the photon is attenuated inside the scintillator, \\
    b. the photon is absorbed in one of the surfaces,\\
    c. the photon reaches PMT, or\\
    d. the photon is still alive and should be followed up.

In the cases a and b the photon is not counted, and should be put
away. If a photon reaches the PMT, the flight time will be saved.
Otherwise, in case d, the photon is not absorbed in the surfaces and
it should be reflected in a new random direction, so we go back to
step 2. For photons reaching PMT, the emission time also should be
added to the time of flight. If for any reason the photon does not
reach the PMT after 200 ns, the program does not follow this photon, 
and goes to the next photon.
\end{enumerate}

To consider the absorption of photons in the walls and the
bottom surface of the light enclosure, a generated uniform random
number $R \in [0,1]$ is compared with the absorption coefficients of
the walls (0.2) and the surface (0.1). If $R$ is less than the
absorption coefficient, this photon is considered as absorbed.
Similarly, for the attenuation of the photons in the scintillator,
another uniform random number should be compared with the
attenuation probability distribution, which is given by $P_{\rm
att.}$:

\begin{equation}
P_{\rm att.} = 1 - e^{-l/\lambda} ;
\end{equation}
where $l$ is the distance traveled by the photon inside the
scintillator, and $\lambda =250\, {\rm cm}$ is the attenuation
length of photons in the NE102A \cite{Knoll-2000}.

\subsection{Results and discussion}

When a charged particle passes through the detector, on average,
41280 photons will be emitted from the scintillator. A small
fraction of them reach the PMT, and most of them will be either absorbed 
on the light enclosure surface or attenuated inside the
scintillator. Table \ref{tab:table1} gives the results of our
extended-code simulation about the fate of the photons emitted from
the scintillator in the light enclosure with a height of 20~cm. The
given values in table 2 are the mean number of photons of all the
cells on the scintillator surface with repeating the simulation 200
times for each cell. According to the number of photons reaching PMT
(485 photons), the photon collection efficiency is $\sim$1\%. This
is consistent with scintillators with WLS fiber readouts, since a
large fraction of the emitted photons will be lost in the process of
photon absorption and re-emission in WLS fibers. For example, the
photon collection efficiency of the scintillation detectors of
GRAPES-3 with WLS fibers obtained by G3Sim code is $\sim$0.4\%
\cite{2012-Mohanty-RSI}.

\begin{table}
\begin{center}
\begin{tabular}{|c|c|c|}
\hline
  Photon statistics & Number & \% \\
\hline
 \textbf{Photons emitted in the Scintillator} & \textbf{41280} & \textbf{100} \\
 Photons reaching PMT & 484 & 1 \\
 Photons attenuated inside the scintillator & 15357 & 37 \\
 Photons absorbed on the bottom surface of the light enclosure & 15349 & 37 \\
 Photons absorbed on the walls of the light enclosure & 10090 & 25 \\
 \hline
\end{tabular}
\caption{\label{tab:table1}Total number of photons emitted from the
scintillator, number of photons reaching PMT, attenuated inside the
scintillator, absorbed on the bottom surface of the light enclosure
and absorbed on the walls of the 20 cm light enclosure.}
\end{center}
\end{table}

It is informative to consider number of collisions of counted
photons (i.e., the photons reaching PMT) with the walls of the light
enclosure. Among photons reaching PMT, some of them do not collide
with the walls of the light enclosure, and the rest collide once, 
twice or more with the walls. Table \ref{tab:table2} presents number
of photons reaching PMT for different numbers of collisions with the
walls.

\begin{table}
\begin{center}
\begin{tabular}{|c|c|c|}
\hline
  No. of collisions & No. of photons & \% \\
\hline
 0 & 75 & 15.5 \\
 1 & 74 & 15.3 \\
 2 & 61 & 12.6 \\
 3 & 51 & 10.5 \\
 4 & 42 & 8.7 \\
 5 & 35 & 7.2 \\
 6 & 28 & 5.8 \\
 7 & 22 & 4.6 \\
 8 & 18 & 3.7 \\
 9 & 15 & 3.1 \\
 10 & 12 & 2.5 \\
 11 & 8 & 1.7 \\
 12 & 6 & 1.2 \\
 13 & 5 & 1.0 \\
 14 & 4 & 0.1 \\
 15 & 4 & 0.1 \\
 \hline
\end{tabular}
\end{center}
\caption{\label{tab:table2} Counted photons with different numbers of
collisions with the walls, averaged over 200 runs, for the cell 3 of
the 20 cm light enclosure.}
\end{table}

More than half of the counted photons collide the walls less than 4
times, and for more than 15 collisions, the number of counted
photons is less than 0.1$\%$ (third column of Table
\ref{tab:table2}). On average 75 photons reach PMT without any
collision with the walls or the bottom surface of the light
enclosure. This number of photons is enough to produce a signal in
PMT. For a typical PMT with quantum efficiency of $\sim$28\%, an
output signal could be produced by $\sim$21 photoelectrons.

\begin{figure}
\begin{center}
\includegraphics*[width=0.7\textwidth]{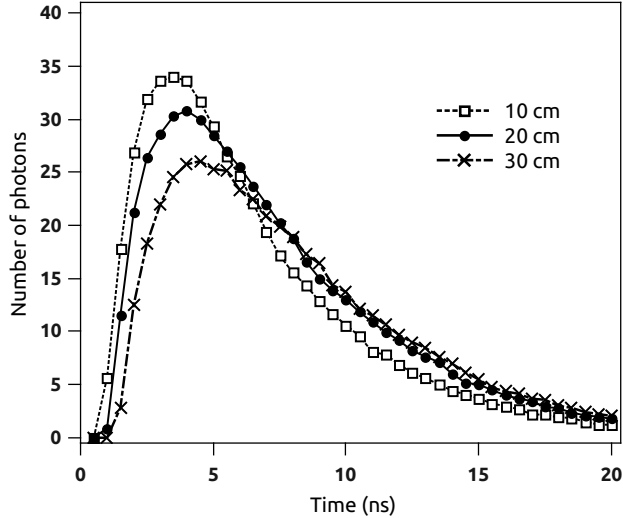}
\caption{\label{fig:3LEspec} Time spectrum of photons reaching PMT
corresponding to cell 3 for 3 light enclosures.}
\end{center}
\end{figure}

The time spectrum of photons, i.e., the number of photons reaching
PMT in 0.5~ns time intervals, gives a very valuable information
about the performance of the detector. The total number of photons
reaching PMT (the area under the spectrum curve), the rise time, the 
FWHM, the fall time, and the shape of the spectrum provide suitable tools 
to compare the detectors with different light enclosure heights. Time
spectra of 3 light enclosure heights corresponding to the third cell
(Fig.~\ref{fig:detector}(c)) are shown in Fig.~\ref{fig:3LEspec}.
Increasing height of the light enclosure has multiple effects on
the spectrum: it shifts the whole spectrum and its maximum toward
the later times; it decreases the height of the spectrum; and it
increases the rise time and FWHM.

\begin{figure}[h]
\begin{center}
\includegraphics*[width=0.7\textwidth]{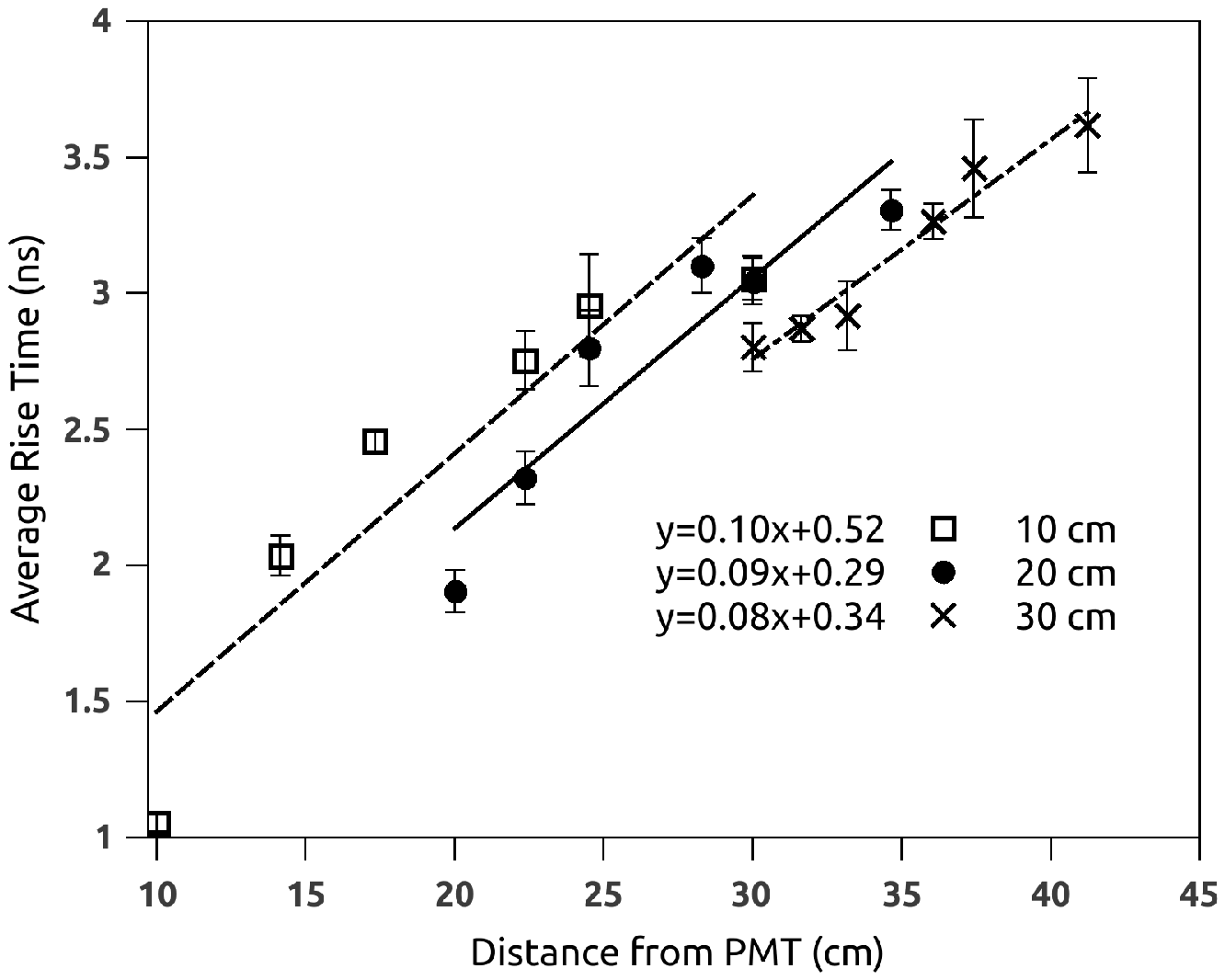}
\caption{\label{fig:risetime} Rise time of 6 cells as a function of
distance between corresponding cell and PMT for 3 light enclosures.
The values are averaged over 200 runs of the program. Fitted lines
and their slopes are given in the graph for 10 cm , 20 cm, and 30 cm
light enclosures. Deviation of rise time points from the fitted
lines could be result of light enclosures geometric asymmetry.}
\end{center}
\end{figure}

The mean rise time of the spectra of 200 runs for each cell of the
detector is plotted as a function of distance from the PMT in
Fig.~\ref{fig:risetime}. The rise time at the center of the detector
is minimum and increases by going toward the corner. The smallest
value for the rise times belong to the 10~cm light enclosure in all
cells, and the corresponding values increase by increasing the light
enclosure height. On the other hand, the slopes of the fitted lines,
in Fig. 6, show that the rise time of the 10 cm light enclosure is
more non-uniform.

Dispersion of the rise times around the average value gives time
resolution of the detector. These values are 0.11~ns, 0.10~ns, and
0.13~ns for 10~cm, 20~cm, and 30~cm light enclosures, respectively.
On the other hand, the time resolution of the 20 cm light enclosure
is obtained by experiment. Error of the time measurement can be
represented by $\sigma_{\rm total}=(\sigma_{\rm
elec.}^{2}+\sigma_{\rm d}^{2})^{0.5}$, where $\sigma_{\rm elec.}$
and $\sigma_{\rm d}$ are, respectively, the errors in time due to the electronic
circuit and the detector (including the light enclosure and the
scintillator). Since $\sigma_{\rm total}$ and $\sigma_{\rm d}$ are
1.85~ns and 0.10~ns, respectively, we obtain $\sigma_{\rm
elec.}=1.847$~ns. The latter error is due to 2 PMTs, 2
discriminators, one TAC and one ADC in the circuit. The timing
errors (by catalogues) of the PMT, the discriminator, the TAC, and 
ADC are 1.27~ns, 0.1~ns, 0.1~ns, and 0.2~ns, respectively. On the 
basis of these errors, the total electronic error is
\begin{equation}
\sigma_{\rm elec.}=(2\sigma_{\rm PMT}^{2}+2\sigma_{\rm Disc}^{2}+\sigma_{\rm TAC}^{2}+\sigma_{\rm ADC}^{2})^{0.5}= 1.82 \, {\rm ns} ;
\end{equation}

Therefore, the time resolution of the detector is $\sigma_{\rm
d}=(\sigma_{\rm total}^{2}-\sigma_{\rm elec.}^{2})^{0.5}=0.31$~ns,
which is comparable with 0.10~ns of simulation.

Regardless of these considerations, there is a general conclusion
which is consistent with the experimental results: increasing the
height of the light enclosure improves the uniformity of the rise time
while decreasing the height diminishes the rise time values.

\begin{figure}
\begin{center}
\includegraphics*[width=0.7\textwidth]{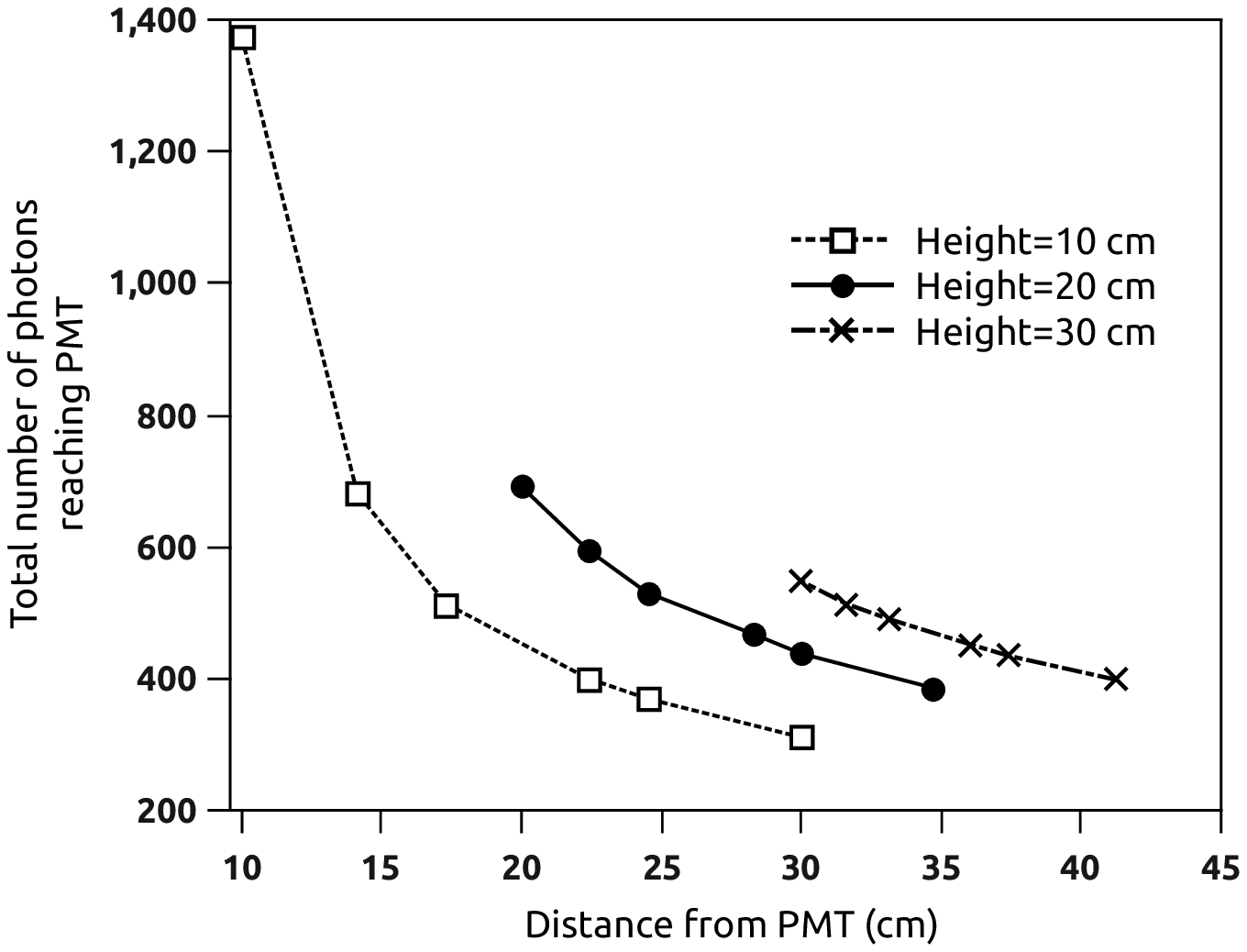}
\caption{\label{fig:NofPhotons} Total number of photons reaching PMT
as a function of distance between corresponding cell and PMT for 3
light enclosures. The numbers are averaged over 200 runs of the
program. Since the error bars are smaller than the data points, they
are omitted.}
\end{center}
\end{figure}

Fig.~\ref{fig:NofPhotons} gives total number of photons reaching
PMT, averaged over 200 runs, as a function of distance between
corresponding cell and PMT. When the number of photons reaching PMT
exceeds a certain value, the efficiency will be 100\%. Below this
value, a little change in the number of photons reaching PMT may
affect the efficiency. For the 10~cm light enclosure, when an
ionizing particle passes through the center of the detector (cell 1
of Fig.~\ref{fig:detector}(c)), more than 1300 photons are counted
in PMT. This is much more than the values related to other cells and
other light enclosure heights. This means that it is reasonable to
estimate the efficiencies of the detectors in terms of the
efficiency of the cell 1 in the 10~cm light enclosure height as
given in Table \ref{tab:table0}. 

We can estimate the response of each detector in terms of photons
reaching PMT. If the threshold height of the signal in the
discriminator is $V_{0}$, for recording an event, the generated signal 
in PMT should have a pulse height more than $V_{0}$. The output 
of PMT is a current, $I_{p}$, while the external signal is a voltage. 
For converting the current output of a PMT into a voltage output, a load 
resistance $R_{L}$ is used. This current for voltage $V_{0}$ is given by 
$I_{0p} = V_{0}/R_{L}$. On the other hand, the current is
\begin{equation}
I_{p} = \dfrac{dQ}{dt} = G\, e\, \dfrac{dN_{p.e.}}{dt} = G\ e \, \eta\dfrac{N_{ph}}{\tau} ;
\end{equation}
where $G$ is the gain of PMT ($\sim$10$^{8}$), $e$ is the electron
charge, $\eta$ is the quantum efficiency of the photocathode
($\sim$28\%), $\dfrac{dN_{p.e.}}{dt}$ is the rate of photoelectrons
produced in the photocathode,
 $N_{ph}$ is the number of photons reaching PMT,
and $\tau$ is the decay time constant of the scintillator
($\sim$2.5~ns).

We can obtain $N_{ph}$ from the current ($I_{p} = V/R_{L}$) and the
other parameters. We denote the number of photons reaching PMT,
corresponding to $I_{0p}$, by $N_{0ph}$. Now, if $N_{ph}>N_{0ph}$,
the signal is recorded, otherwise it is not recorded. Thus, the
response of the detector is estimated.

\subsection{Geant4 cross check}

To cross check our extended-code simulation, an another simulation
was performed using the Geant4 toolkit \cite{Geant4-2006}, only for
the 20~cm light enclosure. To converge the results, we had to modify
our extended-code simulation. As it was mentioned in
{\S\ref{SS3.1}}, in our extended-code simulation, for $dE/dx$ and
$dN/dl$ we used the values $2\, {\rm MeV}/{\rm cm}$ and
$20,000$~photons/cm, respectively. These values are reasonable for
MIP in a low-$Z$ material, but are not reasonable for NE102A.  In
the Geant4 simulation, for the scintillator NE102A, the H/C ratio
(number of the hydrogen atoms to carbon atoms) is considered to  be 
1.105 {\cite{Leory-2004}}. This means that the total number of
photons emitted from each 1 cm of the scintillator, in extended
code, is set to have the same value as that of a 1~GeV electron obtained
by Geant4. To converge the results of the two simulations, we have
to normalize the number of emitted photons.

\begin{figure}
\begin{center}
\includegraphics*[width=0.7\textwidth]{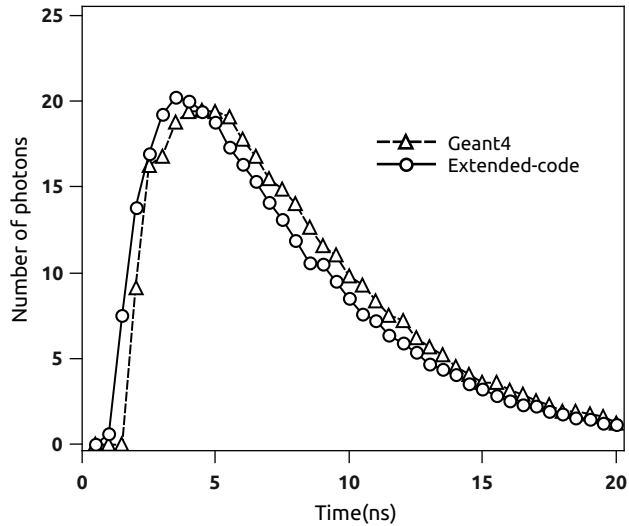}
\caption{\label{fig:geant} Time spectrum of photons: number of
photons reaching PMT in 0.5~ns time intervals, obtained by
Extended-code (300 times running) and Geant4 (200 times running)
simulations corresponding to cell 3 of a 20~cm light enclosure.}
\end{center}
\end{figure}

Fig.~\ref{fig:geant} shows the photon spectrum of the Geant4 and
Extended-code simulations for the cell 3 of a 20~cm light enclosure.
The spectrum of Geant4 is the result of 200 times running of the
code and of extended-code 300 times. There is a good agreement
between the results of the two simulations.

\section{Conclusion}

The experiment in cosmic ray laboratory at Sharif University of Technology 
shows that the 20~cm height is almost an optimum height for the light 
enclosures of the Alborz-I scintillators. This conclusion is confirmed by 
the extended-code and the Geant4 simulations. The simulations show that 
for the detectors of Alborz-I, only 1\% of emitted photons from the 
scintillator can reach the PMT. Since the detectors of Alborz-I will 
be used for timing measurement and not for energy deposition of 
particles, shape and material of the light enclosure do not play 
an important role for these detectors. In order to use the detectors 
for measurement of energy deposition, the uniformity plays an important 
role and using the WLS fiber readouts is a better choice. This is the 
case for the current experiments --- such as new generation detectors of 
GRAPES3 \cite{2009-Gupta-AP}, shielded plastic scintillators of KASCADE
experiment for detection of muons \cite{2003-Ulrich-NIMA}, and
electromagnetic detectors of LHAASO-KM2A \cite{2014-Zhao-CPC} --- which 
use scintillators for detection. Total time resolution of the Alborz-I 
detectors, including the scintillator, the light enclosure, and the PMT, 
is found to be $1.31$~ns. This time resolution fully satisfies our
needs for studying the extensive air showers by the timing methods.

\section{Acknowledgements}
 The authors wish to express their gratitude to all members of Samimi
Cosmic Ray Laboratory (SCRL), specially Saba Mortazavi Moghaddam for
her help in doing experiments, Hadi Hedayati for his help in the
first version of extended-code simulation, Soheila Abdollahi and Mr.
Orami for their help and support. Mohsen Akbari, Ahmad Shariati,
Amir Aghamohammadi and Encieh Erfani are thanked for numerous constructive 
comments. The authors are very grateful for the invaluable and constructive
comments of the anonymous referees.


\end{document}